\documentclass[longauth]{aa} 

\usepackage{graphicx}
\usepackage{gensymb}
\usepackage[normalem]{ulem}
\usepackage{multirow}

\usepackage{txfonts}
\begin{document} 

   \title{The Flux Distribution of Sgr~A*}

   \author{The GRAVITY Collaboration: R.~Abuter\inst{8} \and A.~Amorim\inst{6, 13} \and M~Baub\"{o}ck\inst{1} \and J.B. Berger\inst{5,8} \and H.~Bonnet\inst{8} \and W.~Brandner\inst{3} \and V.~Cardoso\inst{13}, Y.~Cl\'{e}net\inst{2} \and P.T.~de~Zeeuw\inst{11,1}\and J.~Dexter\inst{14, 1} \and A.~Eckart\inst{4,10} \and F.~Eisenhauer\inst{1} \and N.M.~Förster~Schreiber\inst{1} \and P.~Garcia\inst{7, 13} \and F.~Gao\inst{1} \and E.~Gendron\inst{2} \and R.~Genzel\inst{1,12} \and S.~Gillessen\inst{1} \and M.~Habibi\inst{1} \and X.~Haubois\inst{9} \and T.~Henning\inst{3} \and S.~Hippler\inst{3} \and M.~Horrobin\inst{4} \and A.~Jim\'{e}nez-Rosales\inst{1} \and L.~Jochum\inst{9} \and L.~Jocou\inst{5} \and A.~Kaufer\inst{9} \and P.~Kervella\inst{2} \and S.~Lacour\inst{2,8} \and V.~Lapeyr\`{e}re\inst{2} \and J.-B.~Le~Bouquin\inst{5} \and P.~L\'{e}na\inst{2} \and M.~Nowak\inst{17} \and T.~Ott\inst{1} \and T.~Paumard\inst{2} \and K.~Perraut\inst{5} \and G.~Perrin\inst{2} \and O.~Pfuhl\inst{8,1} \and G. Ponti\inst{15, 1} \and G.~Rodriguez~Coira\inst{2} \and J.~Shangguan\inst{1} \and S.~Scheithauer\inst{3} \and J.~Stadler\inst{1} \and O.~Straub\inst{1}, C.~Straubmeier\inst{4} \and E.~Sturm\inst{1} \and L.J.~Tacconi\inst{1} \and F.~Vincent\inst{2} \and S.D.~von~Fellenberg\inst{1} \and I.~Waisberg\inst{16} \and F.~Widmann\inst{1} \and E.~Wieprecht\inst{1} \and E.~Wiezorrek\inst{1} \and  J.~Woillez\inst{8} \and S.~Yazici\inst{1,4} \and G.~Zins\inst{9}}

   \institute{Max Planck Institute for Extraterrestrial Physics, Giessenbachstr. 1, D-85748 Garching bei Muenchen, Germany
              \and
              LESIA, Observatoire de Paris, Universit\'{e} PSL, CNRS, Sorbonne Universit\'{e}, Universit\'{e} de Paris, 5 place Jules Janssen, 92195 Meudon, France 
              \and
              Max-Planck-Institute for Astronomy, K\"{o}nigsstuhl 17, 69117, Heidelberg, Germany
              \and
              1. Physikalisches Institut, Universit\"{a}t zu K\"{o}ln, Z\"{u}lpicher Str. 77, 50937, K\"{o}ln, Germany
              \and
              Univ. Grenoble Alpes, CNRS, IPAG, 38000 Grenoble, France
              \and
              Universidade de Lisboa - Faculdade de Ci\^{e}ncias, Campo Grande, 1749-016 Lisboa, Portugal
              \and
              Faculdade de Engenharia, Universidade do Porto, Rua Dr. Roberto Frias, 4200-465 Porto, Portugal
              \and 
              European Southern Observatory, Karl-Schwarzschild-Str. 2, 85748, Garching, Germany
              \and
              European Southern Observatory, Casilla 19001, Santiago 19, Chile
              \and
              Max-Planck-Institute for Radio Astronomy, Auf dem H\"{u}gel 69, 53121, Bonn, Germany
              \and
              Sterrewacht Leiden, Leiden University, Postbus 9513, 2300 RA Leiden, The Netherlands
              \and
              Departments of Physics and Astronomy, Le Conte Hall, University of California, Berkeley, CA 94720, USA
              \and
              CENTRA - Centro de Astrof\'{\i}sica e Gravita\c{c}\~{a}o, IST, Universidade de Lisboa, 1049-001 Lisboa, Portugal            			  \and
              Department of Astrophysical \& Planetary Sciences, JILA, Duane Physics Bldg., 2000 Colorado Ave, University of Colorado, Boulder, CO 80309
              \and 
              INAF-Osservatorio Astronomico di Brera, Via E. Bianchi 46, I-23807 Merate (LC), Italy
              \and
              Department of Particle Physics \& Astrophysics, Weizmann Institute of Science, Rehovot 76100, Israel
              \and 
              Institute of Astronomy, University of Cambridge, Madingley Road, Cambridge CB3 0HA, UK
              }

   \date{Received September \today; accepted \today}

  \abstract
   {\textbf{The Galactic Center black hole Sagittarius A*} is a variable NIR source that exhibits bright flux excursions called flares. When flux from Sgr~A* is detected, the light curve has been shown to exhibit red noise characteristics and the distribution of flux densities is non-linear, non-Gaussian, and skewed to higher flux densities. However, the low-flux density turnover of the flux distribution is below the sensitivity of current single-aperture telescopes. For this reason, the median NIR flux has only been inferred indirectly from model fitting, but it has not been directly measured. In order to explore the lowest flux ranges, to measure the median flux density, and to test if the previously proposed flux distributions fit the data, we use the unprecedented resolution of the GRAVITY instrument at the VLTI. We obtain light curves using interferometric model fitting and coherent flux measurements. Our light curves are unconfused, overcoming the confusion limit of previous photometric studies. We analyze the light curves using standard statistical methods and obtain the flux distribution. We find that the flux distribution of Sgr~A* turns over at a median flux density of $(1.1\pm0.3)~\mathrm{mJy}$. We measure the percentiles of the flux distribution and use them to constrain the NIR K-band SED. Furthermore, we find that the flux distribution is intrinsically right-skewed to higher flux density in log space. Flux densities below $0.1~\mathrm{mJy}$ are hardly ever observed. In consequence, a single powerlaw or lognormal distribution does not suffice to describe the observed flux distribution in its entirety. However, if one takes into account a power law component at high flux densities, a lognormal distribution can describe the lower end of the observed flux distribution. We confirm the RMS-flux relation for Sgr~A* and find it to be linear for all flux densities in our observation. We conclude that Sgr~A* has two states: the bulk of the emission is generated in a lognormal process with a well-defined median flux density and this quiescent emission is supplemented by sporadic flares that create the observed power law extension of the flux distribution.}
   \maketitle
\section{Introduction\label{section:introduction}}
The supermassive black hole at the Galactic Center, Sagittarius A* (Sgr~A*) is associated with a variable radio/(sub-)mm source, a variable near-infrared (NIR) source, and a continuum source in the X-ray coupled with occasional strong X-ray flares \citep{Genzel2010}.

The NIR counterpart of Sgr~A* is highly variable and not always detected in photometry of ground-based telescopes and space observatories. When the emission is detected, it shows a non-Gaussian flux distribution. The power spectral density (PSD) is best fit with a single power law slope, $\Gamma \sim2$, that breaks into uncorrelated white noise for timescales longer than $\sim250 ~\mathrm{min}$ \citep{Witzel2018}. There is no evidence for quasi-periodic oscillations if the light curve is studied in its entirety \citep{Do2009}; however, individual flares may possess periodic sub-structure \citep{Genzel2003}. The NIR flux distribution has been modeled with a multi-component distribution function, where the fainter flux levels, if detected, are described by a lognormal distribution and the brighter so-called flare states follow a power law tail \citep{Dodds-Eden2009}. However, follow-up studies of the flux distribution have found that a power law tail is not necessary and, instead, a single power law distribution or lognormal distribution suffices to describe the observed distribution of flux densities when temporal correlations are taken into account \citep{Witzel2012, Witzel2018}. By comparing the inferred spectral slope from parallel observations in the NIR K and M band, \cite{Witzel2018} favor a lognormal distribution for both bands. 

Recently, \cite{Do2019} reported a flare of unprecedented brightness (magnitude $\sim12$ in $\mathrm{Ks}$). They find that a flare of this brightness is inconsistent with the long term flux distribution published in \cite{Witzel2018}. They argue that this may indicate that the accretion flow has changed, possibly due to the pericenter passage of the star S2 or the gaseous object G2 \citep{Gillessen2012}. Alternatively, a second mechanism may be needed for the flare state. 

The X-ray flares are correlated with strong NIR flares, but the converse is not true (e.g., \citealt{Dodds-Eden2010, Witzel2012}). While many dedicated multi-wavelength campaigns have been conducted, no clear correlation between either the X-ray or the NIR with the (sub-)mm flux could be established. This is possibly due to  the roughly eight-hour timescale in the sub-mm light curve being on the order of the maximum length of ground-based observations \citep{Dexter2014}.

From a theoretical point of view, the mechanism or mechanisms that generate the NIR flux are not well understood. Because the rise and fall time of flares is on the order of a few minutes, the emitting region is constrained by the light speed to a few Schwarzschild radii $R_s$. Assuming the magnetic field scales as one over the distance with respect to Sgr A*, this constrains the emitting region to be located within $\sim 10 R_s$ of Sgr A* (e.g., \citealt{Barriere2014}). 

The light curve and the slope of bright NIR/X-ray flares have been modeled quantitatively by assuming that a population of electrons is accelerated out of thermal equilibrium into a power law energy distribution. In this model, a cooling break, induced by the frequency-dependent cooling time of synchrotron radiation, explains the NIR and X-ray spectral slopes \citep{Dodds-Eden2010, Li2015, Ponti2017}. However, the mechanism that could explain such an acceleration is not understood in a qualitative fashion. Among several alternatives, previous works have proposed magnetic reconnection as a possible mechanism at work here, drawing upon analogies to solar flares or coronal mass ejections \citep{Yuan2003, YusefZadeh2006}.

Time-dependent simulations attempt to model the accretion flow more qualitatively. The plasma evolution is computed by solving the magnetohydrodynamic (MHD) equations, while accounting for general relativistic (GR) effects in the proximity of the black hole. Such time-dependent simulations have been able to reproduce the typical observational charactaristics of NIR light curves. Realistic light curves have been produced in simulations where the accretion flow is misaligned with the black hole spin \citep{Dexter2014} through lensing of flux tubes \citep{Chan2015}, description of the electron thermodynamics (e.g., \citealt{Ressler2017}, Dexter et al. in prep.), or through the introduction of non-thermal electrons (e.g., \citealt{Ball2016, Mao2016}). However, all of these simulations are limited by the numerical resolution, the volume size of the simulation, and the uncertainty of the initial magnetic field configuration. They have difficulties producing realistic outflows along the poles and cannot produce the observed high X-ray fluxes during flares. 

Recently, the \cite{GRAVITYCollaboration2018} has reported the detection of orbital motions for three bright flares. The three flares exhibit a circular clockwise motion on the sky with typical scales of $150 ~\mathrm{\mu as}$ over a few tens of minutes. This implies a hotspot velocity of around 30 \% of the speed of light. The motion is correlated with an on-sky rotation of the polarization angle with about the same period as the motion. 
Using the relativistic ray tracing code NERO, the GRAVITY Collaboration et al. (submitted) modeled the motions with a hotspot orbiting the black hole, with a roughly face-on inclination of $i\sim 140\degree$. The emitting region is constrained to less than five gravitational radii in diameter. 

In this paper, we build on our previous work on the flux distribution, extending the measurements beyond the detection limit of single-telescope observations using interferometric model fitting and coherent flux measurements. The high sensitivity of GRAVITY pushes the detection limit well beyond the peak of the flux distribution, which allows us to establish an empirical median flux density and variability measures. Through interferometric model fitting, we obtain the un-confused source flux densities of Sgr A* and thereby overcome a fundamental limitation of single telescope observations. Furthermore, we test the paradigm of a single probability distribution for the flux distribution and test different probability density functions (PDFs).

\section{Data\label{section:data}}
There are two independent methods for extracting the flux from our interferometric data. The first method is similar to traditional photometry where we measure the integrated coherent flux. The coherent flux is computed as the flux product of each baseline consisting of a pair of telescopes, normalized by the visibility on this baseline. The coherent flux is blind to incoherent light, that is, speckle noise from bright nearby stars is suppressed. Explicitly, we compute the coherent flux as:
\begin{align}
    \langle F_{coherent}\rangle =  |~\langle A_{vis}\cdot \exp{(-i \phi_{vis})} \rangle_B \cdot \langle F\rangle_B~|,
\end{align}
where $A_{vis}$ is the visibility amplitude, $\phi_{vis}$ is the visibility phase, $F$ is the detector flux and $\langle \circ \rangle_B$ denotes the average over all baselines. To calibrate the flux density, we compute the coherent flux of observations centered on S2. We interpolate the flux in the time gaps between calibration observations using polynomial fits. We use a zeroth order polynomial if there are fewer then three calibrator measurements, a first order polynomial if there are fewer than five calibrator measurements, and a second order polynomial if there are five or more measurements. The coherent flux of S2 is closely correlated with airmass. If extrapolation is necessary, we check that it is reasonable. Explicitly, we checked that the extrapolation does not diverge and that it  resembles the air mass trend. To account for the fact that S2 may not be perfectly centered with respect to the actual fiber position, we calibrate the visibility phase to $0\degree$. 

The second method to measure the flux density uses a model fitting applied to the observed interferometric quantities: the visibility modulus, the visibility squared, and the closure phase. We can model the GRAVITY Galactic Center observations with an interferometric binary consisting of Sgr~A* and the orbiting star S2. According to the van Cittert-Zernicke theorem, the visibility of an image is given by the Fourier transform of the image. In the simple binary case, where S2 and Sgr~A* are modeled as two point sources of a given flux ratio $f$, separated by a certain distance $s = \Vec{B} \cdot \Vec{\delta D}$, the complex visibility is given by:
\begin{equation}
    V(s, f) = \dfrac{1 + fe^{(-2\pi i s)/\lambda}}{1 +f},
\end{equation}
for a given baseline vector $\Vec{B}$, separation vector $\Vec{\delta D}$ and wavelength $\lambda$.  

For real observations, this formula needs to be extended to account for various different parameters such as the source spectral slopes, potentially varying flux ratios for different baselines, pixel response functions, etc. The full derivation of the fitting formula can be found in \citet{ReisWaisberg2019}. 

\begin{figure*}
    \centering
    \includegraphics[width=\textwidth]{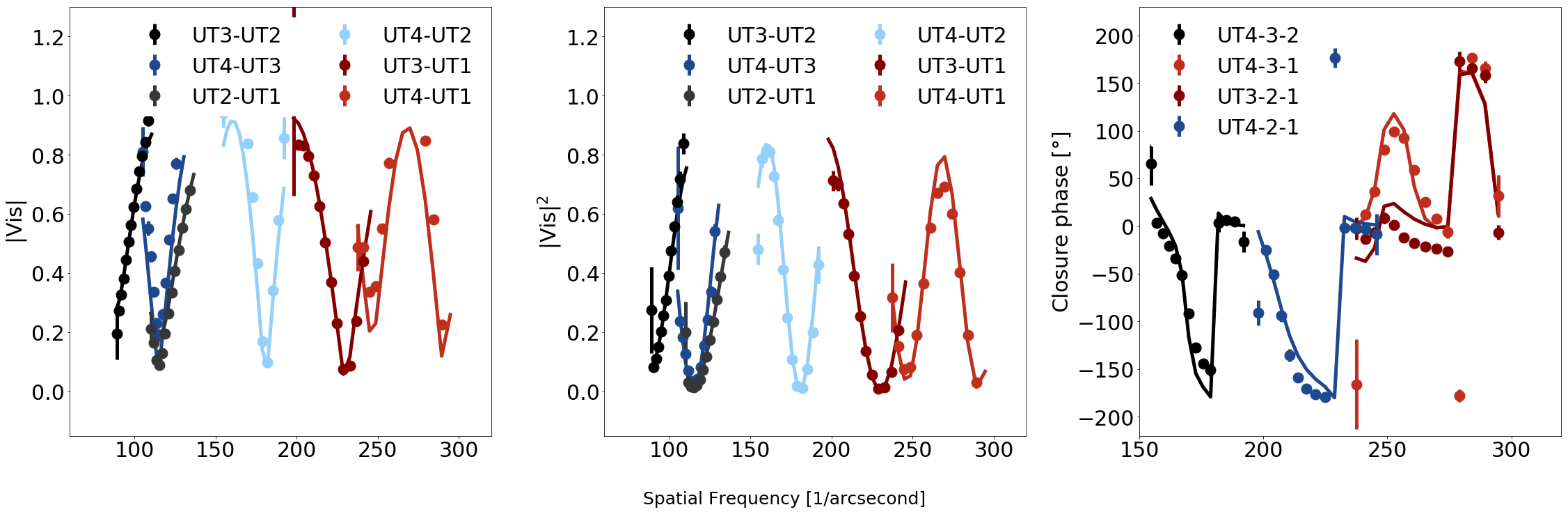}
    \caption{Binary fit to interferometric quantities for the night of July 28, 2018. The three panels show the best-fit binary model to the visibility modulus, the visibility squared, and the closure phase.}
    \label{fig0:ModelFit}
\end{figure*}

Figure~\ref{fig0:ModelFit} shows an example of the full binary model fit to the observed visibilities and closure phases for the night of July 28, 2018. The angular separation of S2 and Sgr~A* was $\sim24.7~\mathrm{mas}$. Moreover, during the observation, a bright flare occurred for which an orbital motion close to the innermost stable orbit was reported in \cite{GRAVITYCollaboration2018}. The flux density of Sgr~A* is $(9.1\pm1.3)~\mathrm{mJy}$.

We chose either the coherent flux or the binary fit to measure the flux density depending on the separation of S2 and Sgr~A*. For 2017 and 2018, strong binary signatures are present in the data. We can therefore make use of the model fitting where the flux ratio is a direct, absolute and un-confused measurement of the flux of Sgr~A*. In 2019, S2  moved to the edge of the interferometric field of view (IFOV), and thus fitting a binary model becomes more difficult. Consequently, we use the integrated coherent flux for 2019. 

The binary flux ratios measure the un-confused flux ratio of the two sources. This assumes that there is no third source hidden near Sgr~A* or S2 within the $\sim (2\times4)~\mathrm{mas}$ interferometric beam of GRAVITY. The 2017 and 2018 light curves of Sgr~A* are unaffected by the contribution of nearby stars overcoming the confusion limit of previous studies. In contrast, the coherent flux includes any possible coherent sources within the IFOV of GRAVITY, corresponding to $\mathrm{FWHM} = \sim 70~\mathrm{mas}$.

In total, our data set comprises 47 nights in 2017 to 2018 and an additional 27 nights in 2019. Prior to 2019, there are $650$ exposures centered on Sgr~A*, totaling to $\sim54.2$ hours. After bad data rejection, $461$ exposures remain totaling to $\sim38.4$ hours. In 2019 there are $324$ observations, out of which $268$ pass the rejection totaling to $26.3$ hours. 

The reduction of the individual exposures is largely unchanged compared to the reduction used in \cite{GRAVITYCollaboration2018a, GRAVITYCollaboration2019}. For the 2017 and 2018 data, we bin the data to five-minute exposures in order to ensure a robust binary fit at the lowest fluxes. For the 2019 data, where we can measure the coherent flux directly, we use sub-exposures binned to 40 seconds. 

We report the flux density at $2.2~\mathrm{\mu m}$. To obtain absolute, dereddened fluxes, we use the extinction coefficient $A_{Ks} = 2.43\pm 0.07$ from \cite{Schodel2010} and \cite{Fritz2011}. We derive the flux density of S2 from the long-term photometry with NACO, yielding $mag_{K_s}=14.12\pm0.076$. Throughout this work we assume that the S2 flux density is constant in time \citep{Habibi2017}. Combining this measurement with the extinction coefficient above, we find the dereddened S2 flux density at $2.18~\mathrm{\mu m}$ to be $15.8\pm 1.5\mathrm{mJy}$. The uncertainty is dominated by the S2 photometry and the extinction uncertainty. Therefore, we neglect the difference in central wavelength of the NACO and GRAVITY bands ($\sim0.02~\mathrm{\mu m}$). Both methods described above have several peculiarities which must be tuned in the data reduction before we can produce a final light curve. The details are given in the next two sections.

\subsection{Tuning of binary flux ratios}
\subsubsection{Outlier contamination}
Bad fits must be flagged and removed. This is critical in the context of estimating the flux distribution: The quality of the fit is a function of the brightness of Sgr~A* and any selection bias may affect the results. In order to minimize the flux dependent bias, we reject data only based on bad observing conditions or data with obvious telescope, facility or instrument problems. These classifiers are flux independent, and therefore the rejection is less critical. However, even such a blind approach may bias the flux distribution if too many bad fits contaminate the light curve. In order to rule out that bad fits significantly contaminate the flux distribution, we  tested different flagging schemes. We find that our results are robust against outlier contamination (see Appendix~\ref{appendix:detection limit}).

\subsubsection{Coupling correction}\label{paragraph:coupling correction}
The flux from the different telescopes is coupled into optical fibers. Therefore, the flux ratio in the binary fits is not only a function of the intrinsic flux ratio but also of the fiber coupling response function. We approximate the fiber coupling response function as a two-dimensional Gaussian in the field, centered on the fiber center (\citealt{Perrin2019x, GRAVITYCollaboration2018a}). We have chosen files in which the fiber is centered on Sgr~A*. Since the distance between S2 and Sgr~A* changes with time, the flux ratio is modulated by this movement. We correct for this modulation by multiplying the flux ratio by the response function. To account for positioning errors of the fiber, we compute the coupling factor using the measured fiber position with respect to S2. 

In 2017, for each telescope, the respective fiber position was often offset by a few mas. This complicates the correction and makes the 2017 light curve sensitive to this effect. In 2018, the fiber positioning was optimized. Furthermore, S2 was closer to Sgr~A* and the binary separation changes less. As a consequence, the fiber coupling correction is less critical in this year. 

\subsection{Tuning of coherent flux measurement}
In 2019, S2 moved to the edge of the $\sim 70~\mathrm{mas}$ IFOV of GRAVITY. However, throughout 2019, the S2 contribution was on the order of a few percent. It is thus necessary to subtract S2's contribution. 

We can model the flux that is coupled into the fiber using the fiber coupling response function used for the binary. However, at the edge of the IFOV, the relation starts to break down. This is especially critical for low fluxes of Sgr~A* for which the contribution of S2 is comparably large. 

To improve the coupling correction, we use the measured binary flux ratio: We fit the flux ratio for each file and divide the fitted flux through the coherent flux of that file. Since S2's contribution is constant during the night, its contribution can be estimated by dividing the binary flux ratio by the coherent flux and averaging this ratio. The median S2 contribution is around $4 \%$ or $\approx 0.4 ~\mathrm{mJy}$. We correct the coherent flux by subtracting each night's median S2 flux from the individual exposures. 

\subsection{Determination of noise\label{section:determination of noise}}
The uncertainty of the binary flux ratios includes the fit uncertainty. However, systematics dominate the errors. 
Consequently, to determine the noise in our light curve, we use two proxy methods. We use the 2019 light curve which has a higher temporal sampling of eight times $40$ seconds per five-minute exposure. We subtract a polynomial fit from each five minute exposure. We determine the noise from the standard deviation of the residuals. The second approach uses the difference between the $0\degree$ and the $90\degree$ polarization for each exposure. We find a consistent power law dependency between the RMS and the flux density for both methods:
\begin{align}
    \sigma(F) = 0.3 \times F^{0.67}.
\end{align}
We find that a single power law slope suffices to describe the noise. We do not find evidence for a flattening of the noise towards lower fluxes, which would indicate a transition to detector read-out noise. The details of this analysis are presented in Appendix \ref{appendix:error model}.

\section{Results\label{section:results}}
\begin{figure*}
    \centering
    \includegraphics[width=\textwidth]{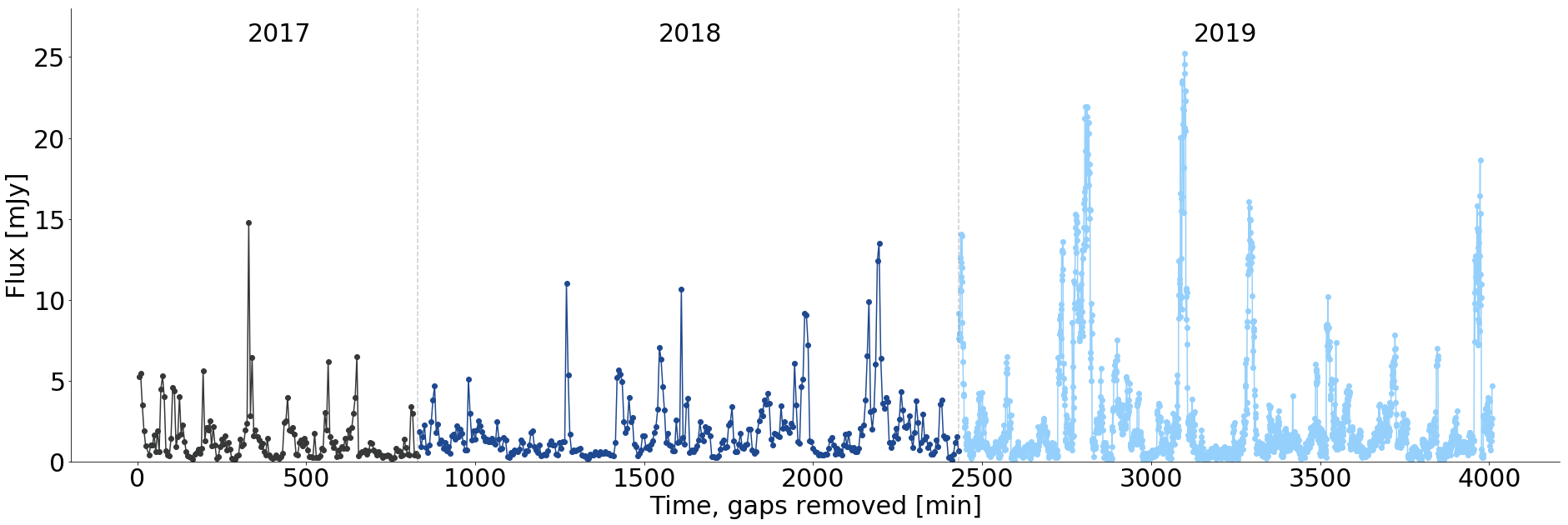}
    \caption{Light curve of Sgr~A* as observed by GRAVITY in the years 2017, 2018, and 2019 with time gaps removed.}
    \label{fig1:light curve}
\end{figure*}

\begin{figure}
    \centering
    \includegraphics[width=0.5\textwidth]{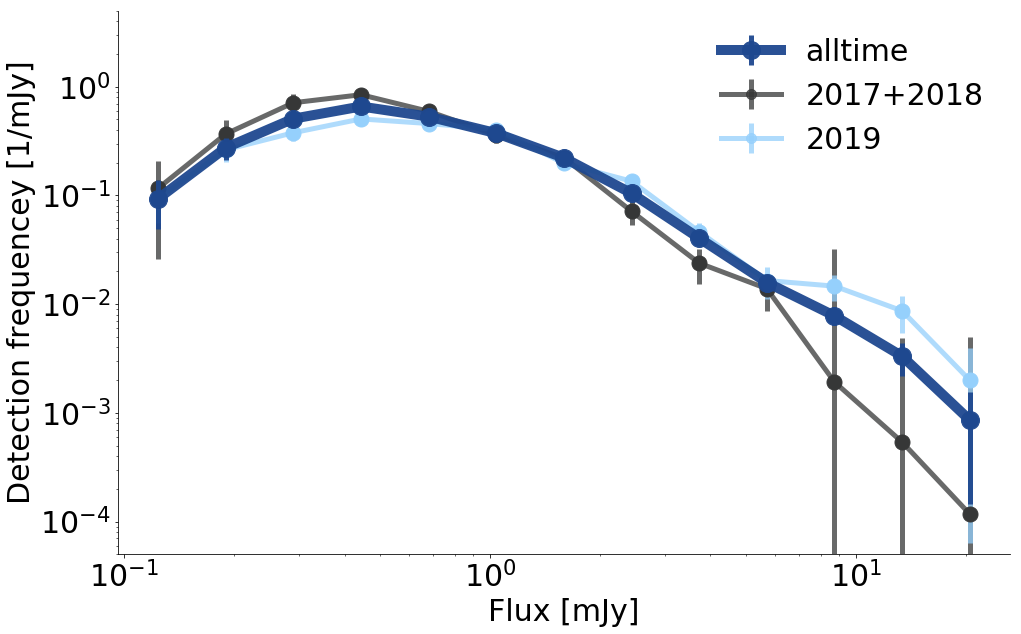}
    \caption{Flux distribution of Sgr~A*: The thick blue line is the flux distribution combining all three observation years. The grey line combines the flux distribution of 2017 and 2018, while the light blue line is the flux distribution of 2019.}
    \label{fig2:flux distribution}
\end{figure}

Figure \ref{fig1:light curve} shows the light curve observed in the years 2017, 2018, and 2019. Figure \ref{fig2:flux distribution} shows the derived flux distribution for the respective years. We choose our histogram bin width and the bin number using Scott's normal reference rule \citep{Scott2015}. This choice is motivated by the fact that the data was well described by a lognormal parameterization in previous studies and our choice of log bins. 

For correlated data, as in our light curve, the $\sigma = 1/\sqrt{N}$ estimator for the bin uncertainty underestimates the errors (e.g., \cite{Vaughan2003}). For this reason, we chose block bootstrapping to estimate the histogram uncertainty. We created $100$ surrogate light curves by copying the original data and dropping 50\% of the observation nights. We redrew the observations nights with replacement from the original data. The choice for blocks of observation nights ensures that the light curve is uncorrelated. We estimate the uncertainty of each histogram bin from the standard deviation of the histogram created from the bootstrapped light curves and quadratically add the $1/\sqrt{N}$ estimate. For 2017 and 2018 we have defined a formal detection significance ratio based on the ratio of significance of a single source model compared to the binary model. If this ratio is below 1, we count the flux density point in the flux density bin where it is observed, but we quadratically add its density contribution to the bin's error.

Sgr~A*'s light curve is correlated and, thus, the flux density histogram is not strictly a measure of the averaged probability density. As a consequence, if Sgr~A* spends less than a correlation time in a certain flux density bin, it is expected that the detection frequency is biased. Because adjacent points in the light curve are correlated, a single high flux density value is likely to be preceded and followed by additional high flux densities. A histogram of such a section of the light curve would overestimate the detection frequency of these high flux densities. Conversely, a section of the light curve containing no high-flux density excursions would lead to an underestimated detection frequency of high flux densities. For observations that last much longer than the correlation timescale, the observed detection frequency converges to the true value. 

While we expect that the block bootstrap captures this effect to some extent. However, it is not clear if it can estimate the errors if more than one physical process in the source is present which may be on or off in different observations. As a consequence, for flux density bins above $3~\mathrm{mJy}$, we conservatively increase the histogram errors by multiplying them by a weight factor. The weight factor is computed by dividing the total observation time for a given flux density bin by a correlation time guess of 120 minutes. While this is shorter than the correlation time estimate in, for instance, \cite{Witzel2018}, it is longer than the usual length of the perceived flares and, consequently, a conservative estimate for a two-process scenario.

It is noteworthy to add that we almost always detect Sgr~A* for all three years. Using our most conservative estimate, we find $17 ~(>3\sigma)$ or $6~(>1\sigma)$ non-detections in 2017 and 2018 (See Appendix \ref{appendix:detection limit} for details). Furthermore, despite the large separation, and consequently the minimal flux coupling of S2, we can almost always fit a binary in 2019. Using reconstructed images we find that Sgr A* is always detected in 2019 (GRAVITY Collaboration in prep.). This illustrates that the flux distribution is right-skewed in log space, and flux densities below $0.1~\mathrm{mJy}$ occur only very infrequently. 

\subsection{The empirical flux distribution \label{section:empircal distribution parameters}}
\begin{figure}
    \centering
    \includegraphics[width=0.5\textwidth]{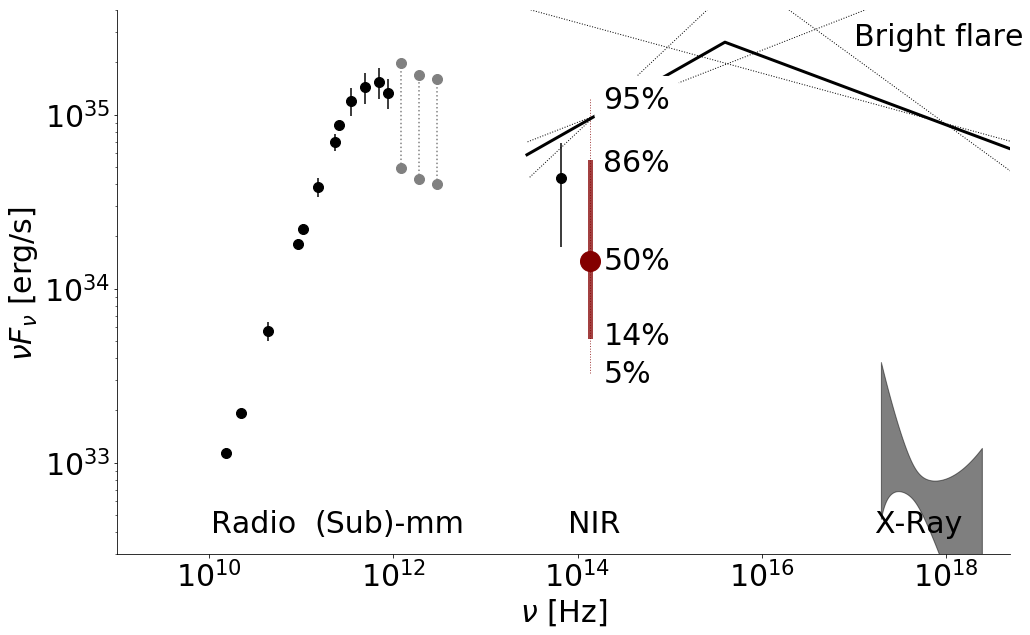}
    \caption{SED of Sgr~A*: the radio and sub-mm data are from \cite{Falcke1998, Bower2015, Brinkerink2015, Liu2016, Bower2019}. The far infrared data is from \cite{Stone2016} and \cite{VonFellenberg2018}. The NIR M-band data is the median flux density inferred from the lognormal model of \cite{Witzel2018}. The NIR K-band data is the GRAVITY flux density: the thick point is the median flux density, and further flux density percentiles are annotated. Also shown are the NIR and X-ray flux density spectrum of a bright simultaneous flare observed by \cite{Ponti2017}, the quiescent X-ray flux density is determined from \cite{Baganoff2003}.}
    \label{fig2:SED}
\end{figure}

Since we almost always have a detection of Sgr~A*, the empirical percentiles can serve as an assumption-free description of the flux distribution. Using the percentiles shown in Table \ref{table1:percentiles}, we updated the SED of Sgr~A* as shown in Figure \ref{fig2:SED}. The uncertainty on the flux density percentile is computed from the difference in the two polarizations, which is believed to be largely instrumental.

Comparing the polarization-averaged flux density percentiles, the 2017 and 2018 50\%, 86\%,  and 95\% percentiles are consistent. Because the 2017 light curve is limited by the fiber coupling correction (see Section~\ref{paragraph:coupling correction}), the low percentiles of 2017 cannot be compared with those of the following years. On the other hand, the low flux density percentiles (5\% and 14\%) of 2018 and 2019 are consistent with each other. The 50\% percentile for 2019 is marginally consistent with its 2017 and 2018 counterpart. 

This is consistent with an unchanged low and median flux distribution in all years covered by GRAVITY observations. 
The high flux density percentiles (86\% and 95\%) of the 2019 data are not consistent with their counter parts from the previous years. This increase in observed flux density is caused by the detection of six bright flares ($F_{SgrA} \sim F_{S2}$) in 2019. The increase in the flux density percentiles is significant with respect to the measurement uncertainty. In the flux distribution we have estimated the bin uncertainty conservatively to account for the correlation in the light curve and the effect of two potential states. In consequence, the flux distribution of 2017 and 2018 is consistent with the 2019 flux distribution.

Table \ref{table1:percentiles} lists separately the percentiles for both $0\degree$ and $90\degree$ polarizations as well as the average. It is not clear if the apparent differences between the two polarization are of physical origin, since the polarization angle is measured with respect to the instrument and not the on-sky orientation. In consequence, the differences may reflect additional systematic uncertainties rather than the intrinsic polarization of the source. 

\begin{table*}
    \centering  
    \caption{Percentiles of the flux distribution: Empirical flux density percentiles of the light curve for the two measured polarizations. The averages reported are the mean of the polarization, the error is computed from the difference of the polarizations. The 5\% and 14\% quantiles of 2017 are affected by instrument systematics and thus are given only for completeness. We note that the polarization angle is with respect to the instrument and is not de-rotated to reflect the on-sky polarization.}             
    \label{table1:percentiles}      
    \begin{tabular}{c|ccccc }
         Percentiles $[\mathrm{mJy}]$: 2017 & 5 \% & 14 \% & 50\% & 86\% & 95\% \\ 
         \hline
         \rule{0pt}{3ex} $0^{\circ}$ Polarization   & \sout{$0.21$} & \sout{$0.29$} & $1.0$ & $2.5$ & $5.0$\\
         $90^{\circ}$ Polarization  & \sout{$0.20$} & \sout{$0.26$} & $0.6$ & $1.7$ & $3.0$\\ 
         \rule{0pt}{3ex} Average & \sout{$0.21\pm0.01$} & \sout{$0.28\pm0.02$} & $0.8\pm0.3$ & $2.1\pm0.6$ & $4.0\pm1.4$ \\
         \rule{0pt}{3ex}
         Percentiles $[\mathrm{mJy}]$: 2018 & 5 \% & 14 \% & 50\% & 86\% & 95\% \\ 
         \hline
         \rule{0pt}{3ex} $0^{\circ}$ Polarization   & $0.35$ & $0.48$ & $1.2$ & $2.8$ & $5.0$\\
         $90^{\circ}$ Polarization  & $0.31$ & $0.43$ & $0.9$ & $2.3$ & $5.1$\\ 
         \rule{0pt}{3ex} Average & $0.33\pm0.03$ & $0.46\pm0.04$ & $1.1\pm0.2$ & $2.6\pm0.4$ & $5.0\pm0.1$ \\
         \rule{0pt}{3ex}
         Percentiles $[\mathrm{mJy}]$: 2019 & 5 \% & 14 \% & 50\% & 86\% & 95\% \\
         \hline
         \rule{0pt}{3ex} $0^{\circ}$ Polarization   & $0.26$ & $0.43$ & $1.2$ & $5.3$ & $12.1$\\
         $90^{\circ}$ Polarization  & $0.34$ & $0.60$ & $1.6$ & $5.6$ & $11.4$\\ 
         \rule{0pt}{3ex} Average &  $0.30\pm0.06$ & $0.5\pm0.12$ & $1.4\pm0.3$ & $5.5\pm0.2$ & $11.8\pm1.3$\\ 
         \hline
         \rule{0pt}{3ex} 2017, 2018 \& 2019 average & $0.28\pm0.07$ & $0.4\pm0.1$ & $1.1\pm0.3$ & $3.4\pm1.7$ & $6.9\pm3.8$ \\
    \end{tabular}
\end{table*}

\subsection{Analytic Distribution Function \label{section:analytic distribution function}}
We fit the flux distribution histogram with several analytic probability density functions (PDFs). These distributions have been selected according to four criteria: non-Gaussianity, right skewedness, historical usage, and physical motivation. These can be grouped into four families of PDFs:

\begin{enumerate}
    \item The lognormal distribution: This distribution is frequently used in active galactic nuclei (AGN) and X-ray binaries: The lognormal distribution results from many unresolved subprocesses which are Gaussian and amplify each other into a single observable. Thus the lognormal distribution results from the product of the Gaussian subprocesses (e.g., \citealt{Uttley2005}).  This model has been applied to Sgr~A* in all past studies of the NIR flux distribution (see, e.g.,\ \citealt{Dodds-Eden2009, Witzel2012, Hora2014, Witzel2018}).
    \item The power law distribution: power law distributions are commonly observed in nature and find a possible explanation in the frame work of self-organized criticality: Self-organized critical systems are systems in which a constant influx of energy breaks down to smaller scales; the power is sometimes associated with the dimensionality of the system or the degrees of freedom \citep{Aschwanden2016}. Such distributions have been discussed in the context of Sgr~A* to explain a possible flare state, the distribution of NIR emission as a whole and the distribution of X-ray flares \citep{Dodds-Eden2009, Witzel2012, Li2015}. 
    \item The family of exponential distributions: Exponential distribution functions such as the Gamma distribution are a generalization of the Poisson process, in which a waiting time between two events is relevant. Such distribution functions have not previously been used to describe the flux distribution of Sgr~A*. However, they are conceptually attractive for accretion flows since the flux density at any time depends on the influx of energy and on the intensity of the flares that had come before. 
    \item Composite distributions: If there are two processes creating the flux, the observed PDF is the convolution of the PDF of each process. Such a two state scenario has been proposed by \cite{Dodds-Eden2009} to overcome the apparent tension of a single lognormal and the high flux flares. In their scenario, the bulk of the emission is created from a lognormal process, and a power law tail is allowed to explain the high flux flares. We adopt this parameterization, but we note that, in principle, many combinations of PDFs could be imagined to explain such a two state scenario.
\end{enumerate}
Before these model PDFs can be fitted to the flux distribution, the effects of measurement noise have to be taken into account. In contrast to single telescope photometric studies, the light curve measured by GRAVITY is unconfused. Prior to 2019, the flux density reported is the direct ratio of S2 and Sgr~A* and is thus unconfused. This assume that there is no third source within the GRAVITY beam (FWHM $\sim (2\times4)~\mathrm{mas}$). 
In 2019 we have measured the integrated coherent flux density in the IFOV, which is blind to the background contribution of bright nearby stars and the galaxy. Furthermore we have subtracted the contribution from S2, which is the closest and brightest star in the IFOV. Using deep images obtained from stacking several observation nights yields an upper limit for the brightness of a potential third source of $\sim 0.3~\mathrm{mJy}$. We therefore assume that Sgr~A* is the only flux contributor in 2019. This assumption is assessed in further detail in appendix \ref{appendix:detection limit}. Consequently, we can model the flux distribution without the assumption of a Gaussian background.

In the presence of observational noise, the intrinsic PDF of Sgr~A* will be affected by the PDF of the noise, that is, the intrinsic distribution function is convolved with the noise distribution. In order to compare our data to a model PDF, we bin the model PDF to match the flux density bins of the observed flux distribution. To address this mathematically, we integrate the noise smoothed model PDF over each histogram bin:
\begin{align}
    P(F) = \dfrac{1}{F_{max} - F_{min}} \int\limits_{F_{min}}^{F_{max}} \int\limits_0^\infty \mathrm{P_{int}}(t,\dots) \cdot \mathcal{N}(\tau,F', \sigma(\tau))~ d\tau ~ dF',
\end{align}
where $F$ is the flux density of the bin center, and $\dots$ substitutes for the intrinsic parameters of the PDF. Since the histogram is normalized to 1, but not all possible flux density states have been observed, we renormalize the observed distribution function.
Assuming the empirical noise relation obtained for the 2019 observations holds for the other years as well, we can model the noise as a Gaussian $\mathcal{N}(F_{obs}, \sigma(F_{obs}))$, where $\sigma= 0.3\times F^{0.67}$.

\subsubsection{Lognormal and power law flux distributions}

\begin{figure}
    \centering
    \includegraphics[width=0.5\textwidth]{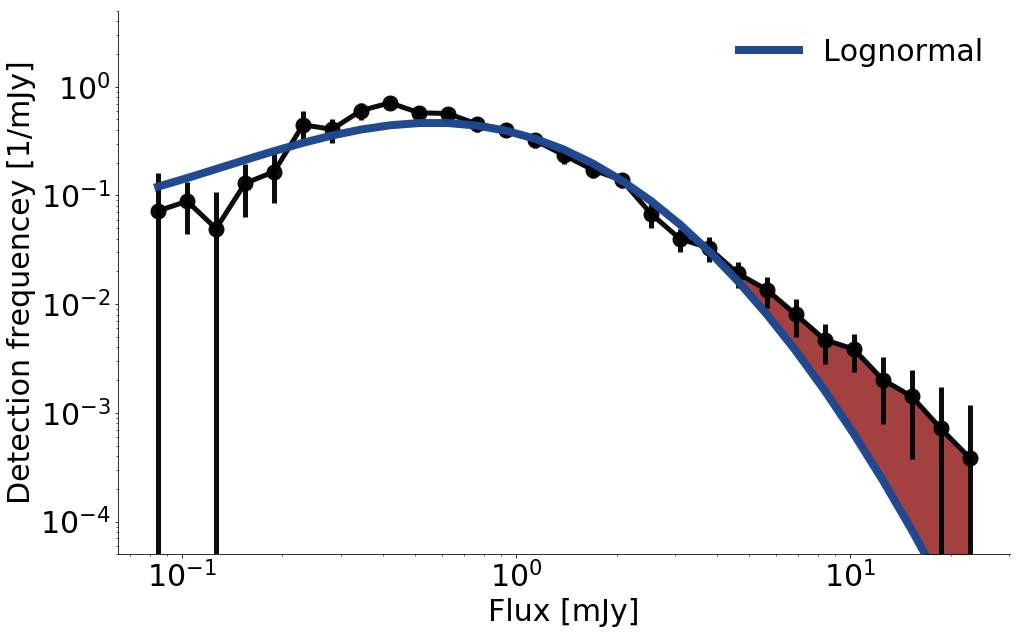}
    \caption{Distribution fits: The observed flux distribution is fitted with a lognormal PDF. The blue line shows the disfavored single lognormal distribution, the red region line indicates the excess flux density compared to the best-fit distribution. We  chose the bin number according to Scott's rule and we chose a logarithmic binning. The error bars are computed from Poisson statistics and a block boot strap: see Section \ref{section:data} for details.}
    \label{fig4:LognormalFit}
\end{figure}

We find that a single lognormal distribution is not sufficient to describe the data. The flux distribution is log-right skewed. Consequently the log-symmetric lognormal distribution cannot fit the tail of the distribution at high flux densities. A lognormal fit to the noise-convoluted distribution function is given in Figure \ref{fig4:LognormalFit}.

Similarly, the detection of the mode of the flux distribution rules out a simplistic power law model with $P(F> F_{min}) = (\alpha+1)/F_{min} \cdot F^{-\alpha}$ and $P(F<F_{min}) = 0$. Nevertheless, the power- law-like tail for flux densities larger than the mode of the flux distribution allows for a variety of models in which high flux densities are described by a power law and the flux densities beyond the mode of the distribution are described by a different parameterization. One such model, the tailed lognormal model proposed by \cite{Dodds-Eden2011} will be discussed in the following subsection. 

\subsubsection{Exponential distribution functions \label{section:exponential distirbutions}}
We test the Gamma distribution and the Weibull distribution as model distributions for Sgr~A*. We find that neither distribution can describe the observed flux distribution. However, when taking their inverse form (i.e., $P(F) = \Gamma(1/F)$) both distribution functions give a good fit to the observed flux distribution. The grey and the dark blue curves in Figure~\ref{fig5:FittingDistributions} show the best-fit inverse Gamma and inverse Weibull PDFs to the flux distribution.

The Gamma function arises from Poisson processes with a distribution of wait times between successive events. This picture makes them initially attractive for modeling the infrared variability as a recurrent flaring process. However, the inverse Gamma function is the same distribution with a random variable corresponding to the reciprocal of the flux density. This quantity can be understood as a timescale with units $[\mathrm{s/erg}]$, that is,\ the time it takes for a certain amount of energy to be released. It is difficult to imagine a physical scenario in which the flux from Sgr~A* can be explained by a succession of events corresponding to an increase in the characteristic emission timescale of the accreting material. We are not aware of any discussion in the literature of such a process. In the absence of a physically motivated model, we do not therefore consider the inverse exponential description of the flux distribution. 




\subsubsection{Composite distribution functions\label{section:composite flux distributions}}

\begin{figure}
    \centering
    \includegraphics[width=0.5\textwidth]{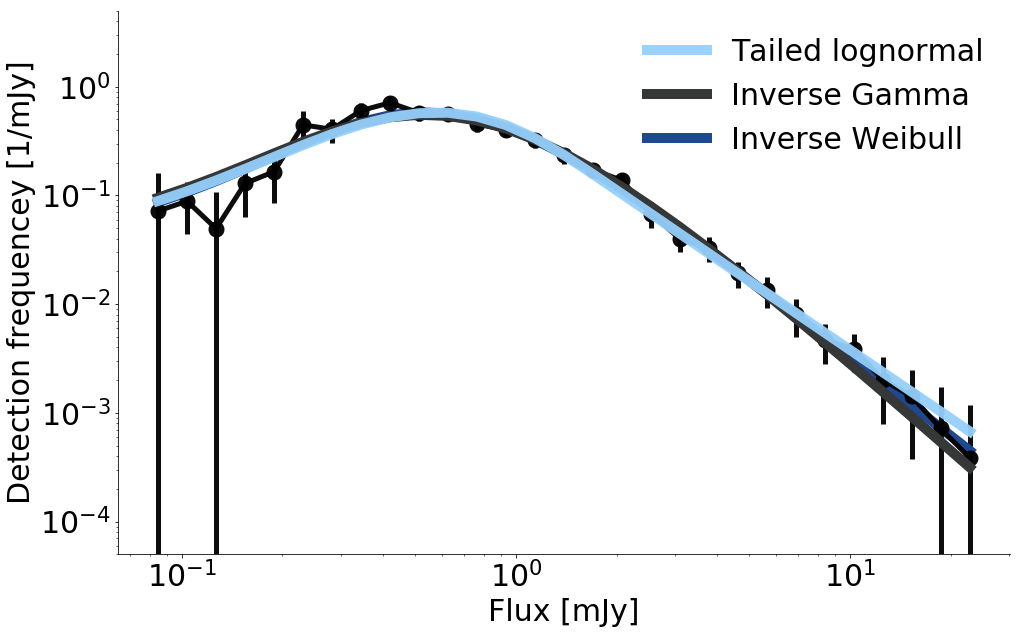}
    \caption{Same as Figure \ref{fig4:LognormalFit}, for distribution functions which describe the observed distribution well.}
    \label{fig5:FittingDistributions}
\end{figure}

We find that a piecewise function consisting of a lognormal distribution joined to a power law tail for flux densities greater than a transition flux density yields a good fit to the flux distribution. Such a distribution function has been proposed by \cite{Dodds-Eden2010} and has been interpreted in the following sense:
The quiescent low flux density states are associated with a lognormal distribution. The lognormal flux distribution is motivated in analogy to the flux distribution of many accreting compact objects such as X-ray binaries or AGN (e.g., \citealt{Uttley2005}). On top of the quiescent phase, there exists a secondary process which creates the flux density tail responsible for the highest flux densities, which coincide with the observed flaring events. The transition flux density marks the flux density at which the observed fluxes are dominated by the secondary process. We fit the distribution function with the parametrization proposed by \cite{Dodds-Eden2010} and find that such a prescription yields a very good fit to the data (see the light blue curve in Figure \ref{fig5:FittingDistributions}). Such a parametrization is useful to illustrate a flux distribution composed of multiple components; however it is not rigorous in a statistical sense: A two-process scenario would be described by the convolution of the individual processes. However, we include it here as a proxy for models in which the flares are described by a separate physical process from the low-flux density state.

\begin{table*}
    \centering  
    \caption{Comparison of a lognormal, tailed lognormal, inverse Gamma and Weibull distribution: Name, functional, best fit values, $\chi^2$, Bayesian Information Criterion (BIC) and the small sample corrected Akaike Information Criterion (AICc). The dimensionless parameters describing flux densities are in mJy.}     
    \begin{tabular}{c|ccccc}
        Distribution name & Functional & best fit values    &   $\chi^2_{red}$      & BIC       & AICc\\
        \hline
        \rule{0pt}{5ex}
        \multirow{2}{*}{LogNormal}  & \multirow{2}{*}{$\dfrac{1}{\sqrt{2\pi}} \dfrac{1}{x\sigma_{ln}} \times \exp{\dfrac{-(\log{x} - \mu_{ln})^2}{(2\sigma_{ln})^2}}$}   &  $\mu_{ln}=(0.08\pm0.04)$ & \multirow{2}{*}{$1.69$}    & \multirow{2}{*}{$52.4$}    & \multirow{2}{*}{$46.1$} \\      
            &   &  $\sigma_{ln}=(0.77\pm0.03)$ &           &           &\\
            
        \rule{0pt}{5ex}
        \multirow{4}{*}{LogNormal + Tail} & 
        \multirow{4}{*}{$\left\{\begin{array}{ll}
                        \mathcal{LN}(x) & x \leq x_{min} \\
                        c\mathcal{LN}(x_{min})F^{-\alpha}/x^{-\alpha}_{min} & x > x_{min}
                        \end{array} \right. $}
                & $\mu_{ln}=(-0.21\pm0.23)~$& \multirow{4}{*}{$0.63$} & \multirow{4}{*}{$29.2$} & \multirow{4}{*}{$17.4$} \\ 
        & &  $\sigma_{ln}=(0.53\pm0.13)~$& & & \\
        & &  $\alpha=(2.08\pm0.12)$& & & \\
        & &  $x_{min}=(1.1\pm 1.9)$& & & \\
        
        \rule{0pt}{5ex}
        \multirow{2}{*}{Inverse Gamma}&\multirow{2}{*}{$\dfrac{\beta^\alpha}{\Gamma (\alpha)x^{(\alpha+1)}} e^{(-\beta/x)}$}& $\alpha=(1.76\pm0.14)$ & \multirow{2}{*}{$0.57$} & \multirow{2}{*}{$22.1$} & \multirow{2}{*}{$15.8$}\\
        & & $\beta=(1.49\pm0.14)$ & & &\\
        
        \rule{0pt}{5ex}
        \multirow{2}{*}{Inverse Weibull}& \multirow{2}{*}{$\beta~\alpha^\beta  x^{(\beta+1)} e^{(-(\alpha/x)^\beta)}$}& $\alpha=(0.78\pm0.03)$ & \multirow{2}{*}{$0.44$}& \multirow{2}{*}{$18.7$} & \multirow{2}{*}{$12.4$}\\
        & &$\beta=(1.41\pm0.07)$ & & &\\
    \end{tabular}
    \label{table2:distribution functions} 
\end{table*}

\subsubsection{Comparison of the distribution fits}
Table \ref{table2:distribution functions} summarizes the least squares distribution fits presented in Figures \ref{fig4:LognormalFit} and \ref{fig5:FittingDistributions}. We assess the four competing models using different standard model comparison formulae. We have disfavored the inverse exponential distribution functions, because we do not find a straight forward physical model. 

In all model comparisons, the visual perception that the lognormal distribution fails to produce the high flux density tail is reflected, despite the larger number of parameters of the tailed lognormal model. For instance, the difference in the small-sample corrected Akaike Information criterion\footnote{For correlated data, the model selection criteria are expected to be over or underestimated. We have ignored this effect.} (AICc) between lognormal and the tailed lognormal model is $\Delta AICc = 46.1 - 17.4 =  28.7$, indicating a very strong evidence in favor of the tailed model.

\subsection{The RMS-flux relation}

\begin{figure*}
    \centering
    \includegraphics[width=1\textwidth]{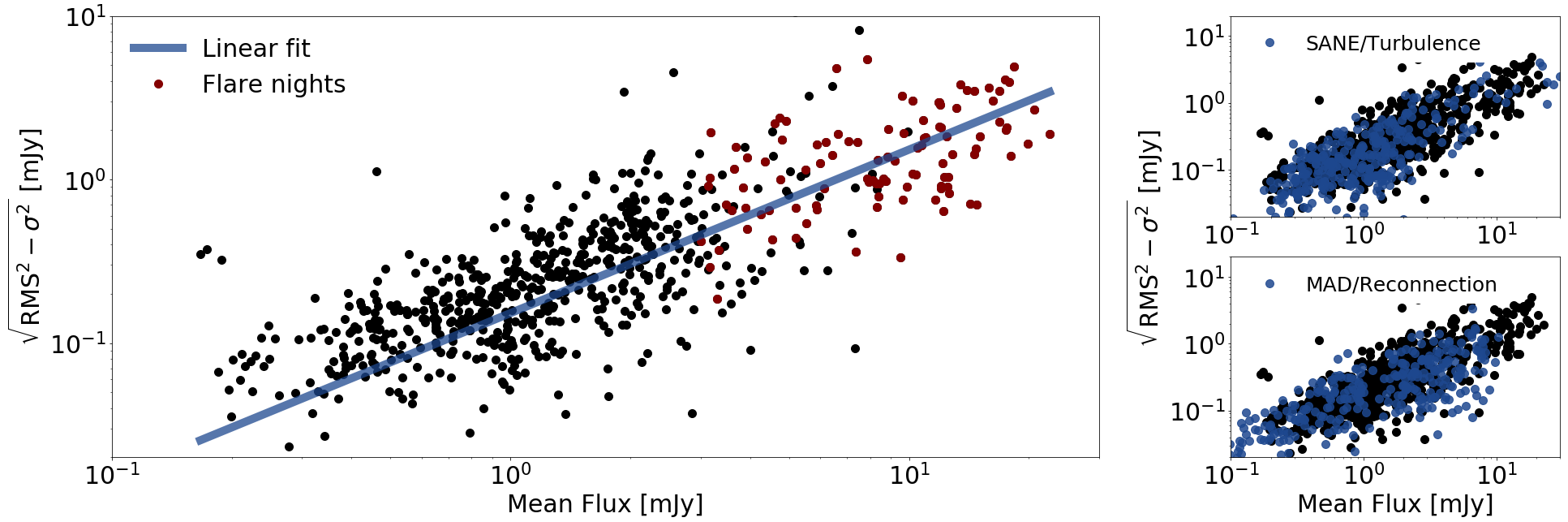}
    \caption{Left: RMS-flux relation of Sgr~A*: The RMS variability of five minute segments of the light curve as a function of the mean flux density in the time bin. The light curve has a data point every 40 seconds, the RMS is computed in the time domain and corrected for the measurement noise $\sigma$. The red points show the relation for mean flux densities above $3~\mathrm{mJy}$ for the six nights with bright flares. The dark blue line is a linear fit, where the RMS values have been weighted using the noise flux density relation determined in section \ref{section:determination of noise}. This accounts for the increasing noise at higher flux densities. Right: Comparison of the observed RMS-flux relation to the relation computed from two GRMHD simulations presented in Dexter et al. 2020 (submitted). The top plot compares the observed relation (black points) to a simulation with a SANE disk (gas pressure dominated) in which electron heating is achieved through a turbulence-like description (dark blue points); bottom plot compares the observations (black points) to MAD disk simulation (magnetic pressure dominated) in which electron heating is achieved through magnetic-reconnection-like description (dark blue points).}
    \label{fig6:rms flux relation}
\end{figure*}

Using the mean and the standard deviation of the 40 second bins of each five minute exposure, we establish the RMS-flux relation for this time scale range. We do not use the integrated power spectrum to determine the RMS, but compute the RMS as $RMS=1/(N-1)\sum^N(x_n-\langle x \rangle)^2$.  The relation is plotted in Figure \ref{fig6:rms flux relation}. In order to correct the RMS for the noise in the measurements, we subtract in quadrature the standard deviation $\sigma$ of the polynomial subtracted light curve to account for the observational errors. 

Every time series generated from a skewed distribution exhibits a relation between the RMS and the mean flux density of a subset of the series (e.g., \citealt{Witzel2012}). Since the RMS of a time series is related to its power spectrum through Parceval's theorem, the RMS-flux relation allows to probe the power spectrum at different mean flux density levels (in the time domain). 

\cite{Vaughan2003} and \cite{Uttley2005} have argued that in the case of a multiplicative lognormal process creating the light curve, the RMS-flux relation is linear on all relevant time scales. \cite{Witzel2012} have reported that Sgr~A* exhibits an RMS-flux relation which is linear to first order. The NACO instrument used by \cite{Witzel2012} is sensitive to timescales on the order of minutes to a few hours.This is too short a time span to effectively probe the variability of Sgr~A* at all relevant time scales; it is shorter than the typical NIR quiescent state correlation time measurements, for instance of $423^{+82}_{-57}$ minutes \citep{Witzel2018}. The same of course applies to GRAVITY, since it is also a ground-based instrument. Consequently, the line of argument used for X-ray binaries, for example, by \cite{Uttley2005} cannot be repeated for Sgr~A* to show a multiplicative process and a lognormal flux distribution. Furthermore, this interpretation has recently been challenged by \cite{Scargle2020}, who argues that both a lognormal and a RMS-flux relation can be created in a shot noise scenario.

Nevertheless, if the power spectrum were different for the higher flux flares, the RMS-flux relation could serve as a tool to disentangle low and high flux density states. We find that the RMS-flux relation is approximately linear. The best-fit linear function has a slope of $1.0 \pm 0.05$ and an abscissa offset of $0.15\pm0.01~[\mathrm{mJy}]$. The red points in Figure \ref{fig6:rms flux relation} are RMS estimates for the six nights for which a bright flare occurred. These points follow the same RMS-flux relation as the low flux density points. Consequently, we find no significant evidence for changed variability during flares. Furthermore we find no flattening of the RMS-flux relation towards the lowest fluxes. This rules out a scenario in which the lowest fluxes are dominated by a second Gaussian source or instrumental limitations. 

The RMS-flux relation can serve as a powerful tool to quantify the variability. It is easily obtained from computing RMS in time domain. This avoids the biases introduced by gaps in the light curve inherent to variability studies in frequency domain. To demonstrate this we compare the observed RMS-flux relation to the relations obtained from two simulations from \cite{Dexter2020} The simulations describe the SED of Sgr~A* well. Both have a duration of roughly 27 hours, assume a black hole spin of $a=0.5$ and an inclination of $i=25\degree$ with respect to the observer. They differ in the ratio of magnetic to gas pressure. For the SANE (Standard And Normal Evolution) simulation the gas pressure dominates. In the MAD (Magnetically Arrested Disk) simulation the magnetic pressure dominates. Furthermore they differ in the description of sub-grid electron heating: The first simulation uses a turbulence-like description and the second simulation uses a description based on magnetic reconnection description. The details of both simulations are described in Dexter et al.\ (2020, submitted). 

We find that both simulations describe the overall variability of Sgr~A* well. The SANE/Turbulence simulation matches the observed variability better, whereas the MAD/Reconnection simulation slightly under-produces the observed mean flux density and variability. This is of course a consequence of the chosen parameters, but demonstrates the use of the RMS-flux relation as an observationally very simple, yet very powerful, tool to constrain models of Sgr~A*.

\section{Discussion}
We find that the flux distribution of Sgr~A* turns over at a flux density of around $0.6~\mathrm{mJy}$ and the empirical median flux density is approximately $1~\mathrm{mJy}$. This bulk of the emission, in the quiescent state, is consistent with remaining through the years of 2017, 2018, and 2019, indicating no immediate effect of the pericenter passage of S2. 

In 2019, we observed six bright flares from Sgr A*. These bright flares cause the flux distribution to extend in a power-law-like fashion for flux densities above $\sim 2 ~\mathrm{mJy}$. We fit the flux distribution with different model PDFs taking into account the effect of observational noise and the binning of the data. Here, the analysis is supported by the fact that our light curves are unconfused. This makes statistical modeling of the background unnecessary. A single power law PDF model is not favored because the flux distribution turns over. It is clear that a bent or broken power law can describe the observed flux distribution. However, without a physically motivated statistical model for the emission of Sgr~A*, such a bent power law does not offer any valuable information. Similarly, we find that distributions of inverse exponential type can describe the log-right skewed flux distribution. However, the inverse form of the distribution function implies a inverse dependence of the flux density to the intrinsic random variable. We associate the inverse flux density to a process-inherent time scale, however we cannot identify such a process. Recently, \cite{Scargle2020} has reviewed a family of flare-like models for astronomical light curves. These models are seemingly able to create arbitrarily shaped flux distributions and linear RMS-flux relations. However, a detailed analysis of the implication of such models is beyond the scope of this paper. 

An alternative to intrinsically log-right-skewed distributions are composite flux distributions. To account for the excess flux density, we allow for an additional power law tail at high flux densities. The tailed lognormal distribution represents a two-state system in which the quiescent emission is created in the first process and the flares cause the power law tail.

We study the variability of the light curve using the RMS-flux relation. We find the RMS-flux relation to be linear for the probed time scale of 40 seconds to five minutes. Intriguingly, we do not observe a change in the RMS-flux relation during the flares.

Based on our finding of a tailed lognormal flux distribution we favor a NIR emission scenario which consists of two components: A quiescent lognormal mechanism that is usually dominant and a separate flare mechanism.
Besides the evidence brought forward in this work and previous works on the flux distribution, there are several additional arguments favoring two distinct NIR states for Sgr~A*.  

\begin{enumerate}
    \item X-ray flares and NIR flares are coupled. The converse is not true (e.g., \citealt{Dodds-Eden2009}).
    \item There is no detectable X-ray quiescent state, which would be clearly associated with the NIR counterpart (eg., \citealt{Genzel2010}).
    \item Strong NIR flares are polarized. The degree of polarization increases with flux density (e.g., \citealt{Eckart2006}).
    \item The spectral index of the flares changes with observed brightness. For flares, the spectral index is $\alpha_{\nu F_{\nu}} \sim 0.5$, but this value decreases to $\alpha_{\nu F_{\nu}}\sim-2$ during the quiescent phase \citep{Gillessen2006}.
    \item \cite{Do2019} detect a $70~\mathrm{mJy}$ flare which is inconsistent with the lognormal flux distribution model of \cite{Witzel2018}, but consistent with a power law tail (G. Witzel, private communication).
    \item Three bright NIR flares have been observed with GRAVITY which show orbital motions. The timescale of the motion is on the same order as the flare duration. Similarly, the observed polarization degree and orientation are correlated with the flare duration and astrometric motion \citep{GRAVITYCollaboration2018}.
\end{enumerate}

\section{Summary}
In this paper, we build on our previous work on the flux distribution into the lowest and highest flux density domains. We detected Sgr~A* in more than $95\%$ of our observations and we conclude that:

\begin{enumerate}
    \item The median flux density $(1.1\pm0.3~\mathrm{mJy})$ as well as the flux density percentiles are robustly measured.
    \item The Sgr~A* SED is constrained by using the measured flux density percentiles. Because we measure flux densities beyond the peak of the flux distribution, we do not have to assume an analytic model for the flux distribution as in previous works (e.g., \citealt{Dodds-Eden2010, Witzel2018}).
    \item The lower percentiles and the median of the flux distribution are stationary within our error estimates and systematic limitations. However, in 2019, we find an increase for the higher percentiles of the light curve. This is due to the observation of six bright flares. 
    \item A single lognormal or power-law-like flux distribution is ruled out. This is because the flux distribution turns over and is log right skewed with a powerl-law-like fall off at flux densities higher than $\sim 2~\mathrm{mJy}$.
    \item The flux distribution is well described by composite distribution functions, such as the tailed lognormal parameterization proposed by \cite{Dodds-Eden2010}.
    \item GRAVITY is the first instrument that allows the study of the variability of the light curve both at fluxes beyond the mode of the flux distribution, as well as the variability of the bright flares. Using the RMS-flux relation, we search for a change in variability during flares. We find a linear RMS-flux relation that holds for both quiescent and flare states.
    \item We conclude that a tailed lognormal PDF describes both the flux distribution and the RMS-flux relation. The two-stated model implied by this parameterization is consistent with all other observational characteristics of the light curve. We thus favor this model over other single-state, right-skewed distribution functions that lack physical motivation.
\end{enumerate}    

Ultimately, the detection of an extreme and unprecedentedly bright flare by \cite{Do2019} and our observations of six additional bright flares in 2019 may indicate that the accretion flow has been altered by the pericenter passage of S2 and/or G2. However, we do not find evidence that the median or mode of the flux distribution has significantly changed in 2019. In consequence, if there are indeed two processes generating the faint quiescent and flaring states, the pericenter passage of S2 or G2 can only have affected the process generating the flares. In light of this constraint, it would be highly interesting to study the sub-mm light curve of Sgr~A*: Since the sub-mm emission is dominated by a population of thermal electrons it measures the particle density and the magnetic properties of the innermost region. Consequently any change in the sub-mm flux distribution in 2019 compared to the previous years may help in understanding the NIR emission scenario.

GRAVITY will continue observing Sgr~A* in the years to come, which will allow for a long-term analysis of the light curve at all flux density levels. This will make it possible to test the long term stationarity of the light curve and possibly yield insights into the changes of the accretion rate. 

\begin{acknowledgements}
SvF, FW, AJ-R \& IW acknowledge support by the Max Planck International Research School. GP is supported by the H2020 ERC Consolidator Grant Hot Milk under grant agreement Nr. 865637. A.A. and P.G. were supported by Funda\c{c}\~{a}o para a Ci\^{e}ncia e a Tecnologia, with grants reference UIDB/00099/2020 and SFRH/BSAB/142940/2018.
\end{acknowledgements}

\appendix
\section{Detection limit \label{appendix:detection limit}}
\subsection{Binary fits}
Data selection plays a crucial role when working with flux ratios obtained by fitting a binary model. We rigorously reject data which has been observed under bad conditions or with instrument malfunctions. In addition, bad fits should be removed from the sample. We must make sure that the MCMC fit has converged. Furthermore, in the case of non- or spurious detections one must ensure that the fit result does not reflect the initial conditions. Most importantly, one must pay special attention that fits are not rejected because of low flux as this skews the resulting flux distribution. 
To ensure that the fitted flux ratios are sound we define four different data selection schemes which we benchmark against each other:

\begin{enumerate}
    \item Manual data rejection: all fit results are visually inspected and the data is qualified according to the quality of the fit and the data. 
    
    \item Astrometric outlier rejection: We calculate the best fit orbit, using all data. We than reject 20 \% of the data that is most outlying, based on the inverse variance weighted distance from orbit position and the fit position. 

    \item Significance of binary rejection: We compare a binary fit to a single point source fit. If the significance binary model is less than $3\sigma$ better than the point source, the data is rejected.
    
    \item No rejection: We use all data points regardless of their apparent quality. 
\end{enumerate}

All of these selection criteria can partially be flux-dependent, even when no data are rejected\footnote{While the SNR of Sgr~A* is flux dependent, the quality of the data is not. For instance, if bad data systematically cause the fits to have artificially high fluxes, those flux bins will be overrepresented in the resulting flux distribution.}. 
To reduce this bias, we redraw the rejected data from the measured accepted data. Such a simple bootstrapping does not take into account the correlation in the light curve, and the data should be re-drawn from self-similar parts of the data (block bootstrapping \citep{Kunsch1989}). However, in our case, the light curve is not long enough to have sampled many high flux states, so a self-similar redrawing from high and low flux states did not alter the results. We thus opt for the simpler bootstrapping approach.  

We find that the manual data rejection (1.) the astrometric outlier rejection (2.) and no rejection yield (3.) consistent results. Only the significance of binary rejection (4.) deviates from the other rejection schemes. In the first three cases, the flux distribution of the rejected data closely follows that of the accepted data. In contrast, the significance of binary rejection scheme shows a strong correlation with the flux and we exclude this scheme. We conclude that the data rejection is mostly unbiased for the manual rejection, the outlier rejection and no rejection. We thus use the simplest scheme, with no rejection, to derive our results.

To asses the detection limit we use different tools. First the convergence of the MCMC chains for the binary fits was checked and proper convergences was ensured. 
We visually checked that binary features are detectable in the visibility amplitude, the squared visibilities and the closure phases. 

To obtain qualitative criteria if Sgr~A* is detected, we explicitly checked all files with a measured flux density below $0.3~\mathrm{mJy}$. We compared the fitted position with the theoretical position based on the orbit. We find that the fitted positions derived at low fluxes do not perform differently from the observations with higher fluxes.

A third qualifier is significance of a binary against a single source model. For the first polarization, we find twelve exposures with a significance below $3 \sigma$ and five exposures with a significance below $1 \sigma$. For the second polarization, $17$ exposures fall below a significance of $3 \sigma,$ and six files below $1 \sigma$. Notably, only one file from 2018 shows a significance below $3 \sigma$. 

We further investigate the files with low fluxes in order to ensure that the binary signal that is observed stems from Sgr~A* and not from another source within the IFOV. We do this by simplifying the fitting procedure and only allow for the binary flux, a visibility scaling and the background flux. We keep the binary separation fixed and run fits over a finely sampled grid with 10000 points $\pm10 ~\mathrm{mas}$ around the best fit position. Because this is computationally expensive we only check one file per year. Since it is expected that a transient background source is visible at least for a few months, this is enough to ensure that the measured signal is not caused by a transient background source.

We find that in both the 2017 and 2018 tested cases, there is significant flux at the separation of Sgr~A* and S2. While there are several degenerate solutions with similar intensity, the fit at the SgrA*--S2 separation has the lowest $\chi^2$. Furthermore, because S2 and Sgr~A*'s positions are very well determined from the orbit fitting (and the other, brighter measurements in the respective nights), we ascribe the degenerate positions at other separations to side lobes of the beam and argue that the S2--SgrA* binary dominates the fit result. 

\section{Flux error model}\label{appendix:error model}
In order to model the flux distribution with an analytic PDF model the observational uncertainties need to be taken into account. Noise has a smoothing effect on the flux distribution. Each measurement is uncertain with a given probability distribution, and when creating a histogram of the data, the measurements may fall into a wrong bin with a probability governed by the error PDF. In this paper we assume the noise to be normally distributed, with a flux-dependent standard deviation. Unlike in similar photometric studies, our noise analysis is limited by the number of observables: The only two observables which are readily available are S2 and Sgr~A*. S2 can only be used to estimate the uncertainty at very high fluxes and it is, thus, of limited use. Furthermore we found the formal fit uncertainties to be poor estimators of the uncertainty. Consequently, we use two empirical approaches to determine the uncertainty. In a first approach, we use the difference between the two polarizations to estimate uncertainty. In the second approach, we assume that the intrinsic light curve is a smooth function, which we can fit with a low order polynomial. We estimate the uncertainties by measuring the standard deviation of the residuals, after subtracting the best fit polynomial. For the second approach we found that polynomials of fourth and fifth order are sufficiently flexible to describe the light curve. 

\begin{figure}
    \centering
    \includegraphics[width=0.5\textwidth]{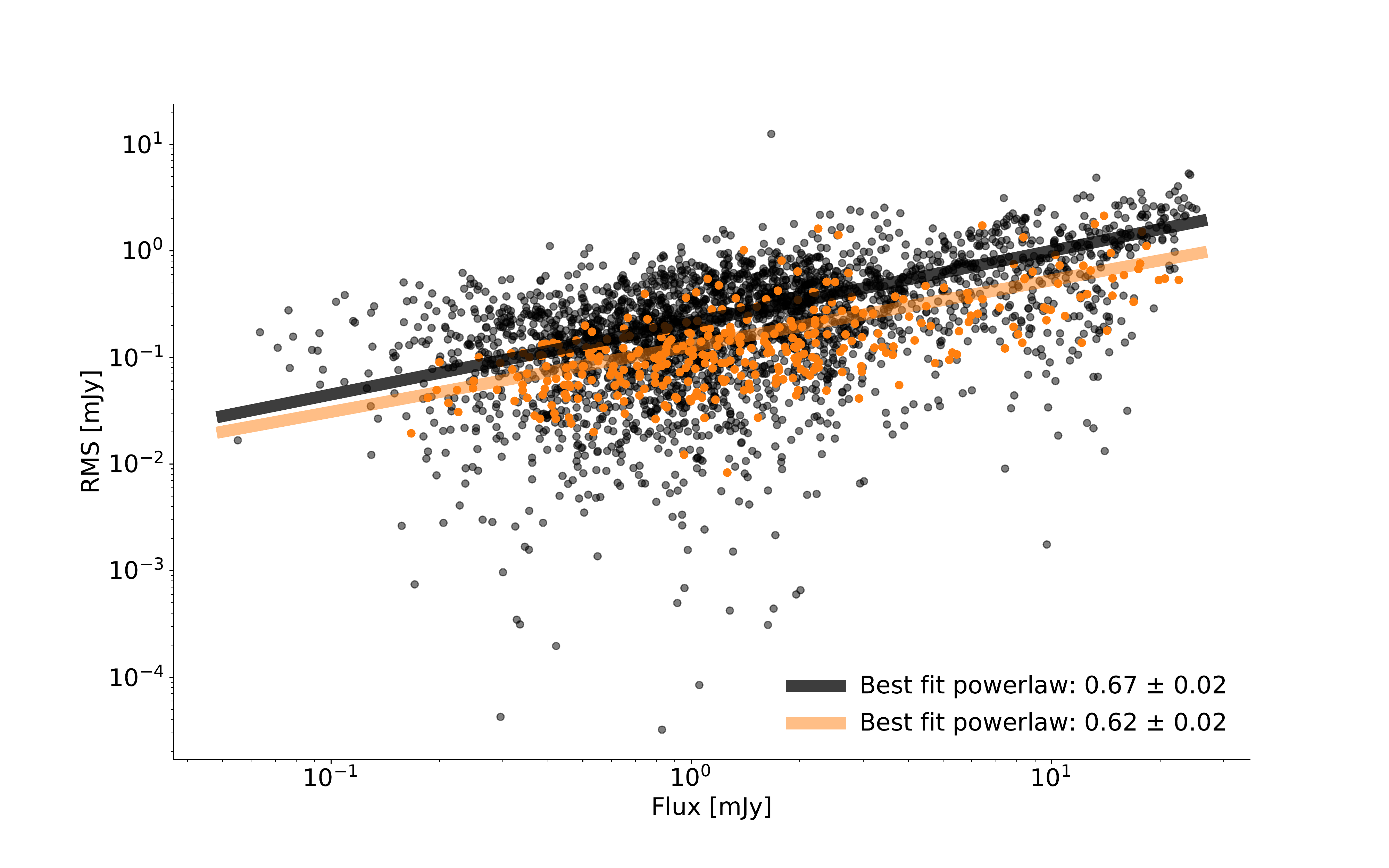}
    \caption{Noise as a function of flux: The RMS is determined from the differences of the two measured polarizations (black points) and the residuals of fourth order polynomial subtracted light curve of Sgr~A*. Both relations can be described by single power law functions, for which we plot the best fitting realization.}
    \label{fig1:noise estimation}
\end{figure}

Both approaches are limited. In the first case, only two measurements determine the uncertainty, and the intrinsic polarization of Sgr~A* inflate the measured uncertainty. In the second case, the assumption of smoothness is imposed, and the order of the polynomial can not be rigorously quantified. However, both measurements quantitatively agree with one another in that the RMS scatter is described by a single power law of $\sigma = 0.3 \times F^{0.6}$, see Figure \ref{fig1:noise estimation}. The exponent of $0.6$ is consistent with the power law description used in photometric studies of the light curve and is consistent with a photon noise origin (e.g., \citealt{Dodds-Eden2011}, \cite{Fritz2011}). 

\bibliography{the_flux_distributionof_sgrA.bib}

\begin{thebibliography}{46}
\expandafter\ifx\csname natexlab\endcsname\relax\def\natexlab#1{#1}\fi

\bibitem[{Yus(2006)}]{YusefZadeh2006}
 2006, The Astrophysical Journal, 650, 189

\bibitem[{Aschwanden {et~al.}(2016)Aschwanden, Crosby, Dimitropoulou,
  Georgoulis, Hergarten, McAteer, Milovanov, Mineshige, Morales, Nishizuka,
  Pruessner, Sanchez, Sharma, Strugarek, \& Uritsky}]{Aschwanden2016}
Aschwanden, M.~J., Crosby, N.~B., Dimitropoulou, M., {et~al.} 2016, Space
  Science Reviews, 198, 47

\bibitem[{Baganoff {et~al.}(2003)Baganoff, Maeda, Morris, Bautz, Brandt, Cui,
  Doty, Feigelson, Garmire, Pravdo, Ricker, \& Townsley}]{Baganoff2003}
Baganoff, F.~K., Maeda, Y., Morris, M., {et~al.} 2003, The Astrophysical
  Journal, 591, 891

\bibitem[{Ball {et~al.}(2016)Ball, {\"{O}}zel, Psaltis, \& Chan}]{Ball2016}
Ball, D., {\"{O}}zel, F., Psaltis, D., \& Chan, C.-k. 2016, The Astrophysical
  Journal, 826, 77

\bibitem[{Barri{\`{e}}re {et~al.}(2014)Barri{\`{e}}re, Tomsick, Baganoff,
  Boggs, Christensen, Craig, Dexter, Grefenstette, Hailey, Harrison, Madsen,
  Mori, Stern, Zhang, Zhang, \& Zoglauer}]{Barriere2014}
Barri{\`{e}}re, N.~M., Tomsick, J.~A., Baganoff, F.~K., {et~al.} 2014, The
  Astrophysical Journal, 786, 46

\bibitem[{Bower {et~al.}(2019)Bower, Dexter, Asada, Brinkerink, Falcke, Ho,
  Inoue, Markoff, Marrone, Matsushita, Moscibrodzka, Nakamura, Peck, \&
  Rao}]{Bower2019}
Bower, G.~C., Dexter, J., Asada, K., {et~al.} 2019, The Astrophysical Journal,
  881, L2

\bibitem[{Bower {et~al.}(2015)Bower, Markoff, Dexter, Gurwell, Moran,
  Brunthaler, Falcke, Fragile, Maitra, Marrone, Peck, Rushton, \&
  Wright}]{Bower2015}
Bower, G.~C., Markoff, S., Dexter, J., {et~al.} 2015, The Astrophysical
  Journal, 802, 69

\bibitem[{Brinkerink {et~al.}(2015)Brinkerink, Falcke, Law, Barkats, Bower,
  Brunthaler, Gammie, {Violette Impellizzeri}, Markoff, Menten, Moscibrodzka,
  Peck, Rushton, Schaaf, \& Wright}]{Brinkerink2015}
Brinkerink, C.~D., Falcke, H., Law, C.~J., {et~al.} 2015, Astronomy and
  Astrophysics, 576, 41

\bibitem[{Chan {et~al.}(2015)Chan, Psaltis, {\"{O}}zel, Medeiros, Marrone,
  Sadowski, \& Narayan}]{Chan2015}
Chan, C.~K., Psaltis, D., {\"{O}}zel, F., {et~al.} 2015, The Astrophysical
  Journal, 812, 103

\bibitem[{Dexter {et~al.}(2020)Dexter, Jiménez-Rosales, Ressler, Tchekhovskoy,
  Bauböck, de~Zeeuw, Eisenhauer, von Fellenberg, Gao, Genzel, Gillessen,
  Habibi, Ott, Stadler, Straub, \& Widmann}]{Dexter2020}
Dexter, J., Jiménez-Rosales, A., Ressler, S.~M., {et~al.} 2020, Monthly
  Notices of the Royal Astronomical Society, staa922

\bibitem[{Dexter {et~al.}(2014)Dexter, Kelly, Bower, Marrone, Stone, \&
  Plambeck}]{Dexter2014}
Dexter, J., Kelly, B., Bower, G.~C., {et~al.} 2014, Monthly Notices of the
  Royal Astronomical Society, 442, 2797

\bibitem[{Do {et~al.}(2009)Do, Ghez, Morris, Lu, Matthews, Yelda, \&
  Larkin}]{Do2009}
Do, T., Ghez, A.~M., Morris, M.~R., {et~al.} 2009, The Astrophysical Journal,
  703, 1323

\bibitem[{Do {et~al.}(2019)Do, Witzel, Gautam, Chen, Ghez, Morris, Becklin,
  Ciurlo, Hosek, Martinez, Matthews, Sakai, \& Sch{\"{o}}del}]{Do2019}
Do, T., Witzel, G., Gautam, A.~K., {et~al.} 2019, The Astrophysical Journal,
  882, L27

\bibitem[{Dodds-Eden {et~al.}(2011)Dodds-Eden, Gillessen, Fritz, Eisenhauer,
  Trippe, Genzel, Ott, Bartko, Pfuhl, Bower, Goldwurm, Porquet, Trap, \&
  2}]{Dodds-Eden2011}
Dodds-Eden, K., Gillessen, S., Fritz, T.~K., {et~al.} 2011, The Astrophysical
  Journal, 728, 37

\bibitem[{Dodds-Eden {et~al.}(2009)Dodds-Eden, Porquet, Trap, Quataert,
  Haubois, Gillessen, Grosso, Pantin, Falcke, Rouan, Genzel, Hasinger,
  Goldwurm, Yusef-Zadeh, Clenet, Trippe, Lagage, Bartko, Eisenhauer, Ott,
  Paumard, Perrin, Yuan, Fritz, \& Mascetti}]{Dodds-Eden2009}
Dodds-Eden, K., Porquet, D., Trap, G., {et~al.} 2009, The Astrophysical
  Journal, 698, 676

\bibitem[{Dodds-Eden {et~al.}(2010)Dodds-Eden, Sharma, Quataert, Genzel,
  Gillessen, Eisenhauer, \& Porquet}]{Dodds-Eden2010}
Dodds-Eden, K., Sharma, P., Quataert, E., {et~al.} 2010, The Astrophysical
  Journal, 725, 450

\bibitem[{Eckart {et~al.}(2006)Eckart, Sch{\"{o}}del, Meyer, Trippe, Ott, \&
  Genzel}]{Eckart2006}
Eckart, A., Sch{\"{o}}del, R., Meyer, L., {et~al.} 2006, Astronomy and
  Astrophysics, 455, 1

\bibitem[{Falcke {et~al.}(1998)Falcke, Goss, Matsuo, Teuben, Zhao, \&
  Zylka}]{Falcke1998}
Falcke, H., Goss, W.~M., Matsuo, H., {et~al.} 1998, {The Simultaneous Spectrum
  of Sagittarius A* from 20 Centimeters to 1 Millimeter and the Nature of the
  Millimeter Excess}, Tech. Rep.~2

\bibitem[{Fritz {et~al.}(2011)Fritz, Gillessen, Dodds-Eden, Lutz, Genzel, Raab,
  Ott, Pfuhl, Eisenhauer, \& Yusef-Zadeh}]{Fritz2011}
Fritz, T.~K., Gillessen, S., Dodds-Eden, K., {et~al.} 2011, The Astrophysical
  Journal, 737, 73

\bibitem[{Genzel {et~al.}(2010)Genzel, Eisenhauer, \& Gillessen}]{Genzel2010}
Genzel, R., Eisenhauer, F., \& Gillessen, S. 2010, Reviews of Modern Physics,
  82

\bibitem[{Genzel {et~al.}(2003)Genzel, Sch{\"{o}}del, Ott, Eckart, Alexander,
  Lacombe, Rouan, \& Aschenbach}]{Genzel2003}
Genzel, R., Sch{\"{o}}del, R., Ott, T., {et~al.} 2003, Nature, 425, 934

\bibitem[{Gillessen {et~al.}(2006)Gillessen, Eisenhauer, Quataert, Genzel,
  Paumard, Trippe, Ott, Abuter, Eckart, Lagage, Lehnert, Tacconi, \&
  Martins}]{Gillessen2006}
Gillessen, S., Eisenhauer, F., Quataert, E., {et~al.} 2006, The Astrophysical
  Journal, 640, 163

\bibitem[{Gillessen {et~al.}(2012)Gillessen, Genzel, Fritz, Quataert, Alig,
  Burkert, Cuadra, Eisenhauer, Pfuhl, Dodds-Eden, Gammie, \&
  Ott}]{Gillessen2012}
Gillessen, S., Genzel, R., Fritz, T.~K., {et~al.} 2012, Nature, 481, 51

\bibitem[{{GRAVITY Collaboration} {et~al.}(2018{\natexlab{a}}){GRAVITY
  Collaboration}, Abuter, Amorim, Anugu, Baub{\"{o}}ck, Benisty, Berger, Blind,
  Bonnet, Brandner, Buron, Collin, Chapron, Cl{\'{e}}net, du~Foresto, de~Zeeuw,
  Deen, Delplancke-Str{\"{o}}bele, Dembet, Dexter, Duvert, Eckart, Eisenhauer,
  Finger, Schreiber, F{\'{e}}dou, Garcia, Lopez, Gao, Gendron, Genzel,
  Gillessen, Gordo, Habibi, Haubois, Haug, Hau{\ss}mann, Henning, Hippler,
  Horrobin, Hubert, Hubin, Rosales, Jochum, Jocou, Kaufer, Kellner, Kendrew,
  Kervella, Kok, Kulas, Lacour, Lapeyr{\`{e}}re, Lazareff, {Le Bouquin},
  L{\'{e}}na, Lippa, Lenzen, M{\'{e}}rand, M{\"{u}}ler, Neumann, Ott, Palanca,
  Paumard, Pasquini, Perraut, Perrin, Pfuhl, Plewa, Rabien, Ram{\'{i}}rez,
  Ramos, Rau, Rodr{\'{i}}guez-Coira, Rohloff, Rousset, Sanchez-Bermudez,
  Scheithauer, Sch{\"{o}}ller, Schuler, Spyromilio, Straub, Straubmeier, Sturm,
  Tacconi, Tristram, Vincent, von Fellenberg, Wank, Waisberg, Widmann,
  Wieprecht, Wiest, Wiezorrek, Woillez, Yazici, Ziegler, \&
  Zins}]{GRAVITYCollaboration2018a}
{GRAVITY Collaboration}, Abuter, R., Amorim, A., {et~al.} 2018{\natexlab{a}},
  Astronomy {\&} Astrophysics, 615, L15

\bibitem[{{GRAVITY Collaboration} {et~al.}(2018{\natexlab{b}}){GRAVITY
  Collaboration}, Abuter, Amorim, Baub{\"{o}}ck, Berger, Bonnet, Brandner,
  Cl{\'{e}}net, {Coud{\'{e}} du Foresto}, de~Zeeuw, Deen, Dexter, Duvert,
  Eckart, Eisenhauer, {F{\"{o}}rster Schreiber}, Garcia, Gao, Gendron, Genzel,
  Gillessen, Guajardo, Habibi, Haubois, Henning, Hippler, Horrobin, Huber,
  Jim{\'{e}}nez-Rosales, Jocou, Kervella, Lacour, Lapeyr{\`{e}}re, Lazareff,
  {Le Bouquin}, L{\'{e}}na, Lippa, Ott, Panduro, Paumard, Perraut, Perrin,
  Pfuhl, Plewa, Rabien, Rodr{\'{i}}guez-Coira, Rousset, Sternberg, Straub,
  Straubmeier, Sturm, Tacconi, Vincent, von Fellenberg, Waisberg, Widmann,
  Wieprecht, Wiezorrek, Woillez, \& Yazici}]{GRAVITYCollaboration2018}
{GRAVITY Collaboration}, Abuter, R., Amorim, A., {et~al.} 2018{\natexlab{b}},
  Astronomy {\&} Astrophysics, 618, L10

\bibitem[{{GRAVITY Collaboration} {et~al.}(2019){GRAVITY Collaboration},
  Abuter, Amorim, Baub{\"{o}}ck, Berger, Bonnet, Brandner, Cl{\'{e}}net,
  {Coud{\'{e}} du Foresto}, de~Zeeuw, Dexter, Duvert, Eckart, Eisenhauer,
  {F{\"{o}}rster Schreiber}, Garcia, Gao, Gendron, Genzel, Gerhard, Gillessen,
  Habibi, Haubois, Henning, Hippler, Horrobin, Jim{\'{e}}nez-Rosales, Jocou,
  Kervella, Lacour, Lapeyr{\`{e}}re, {Le Bouquin}, L{\'{e}}na, Ott, Paumard,
  Perraut, Perrin, Pfuhl, Rabien, {Rodriguez Coira}, Rousset, Scheithauer,
  Sternberg, Straub, Straubmeier, Sturm, Tacconi, Vincent, von Fellenberg,
  Waisberg, Widmann, Wieprecht, Wiezorrek, Woillez, \&
  Yazici}]{GRAVITYCollaboration2019}
{GRAVITY Collaboration}, Abuter, R., Amorim, A., {et~al.} 2019, Astronomy {\&}
  Astrophysics, 625, L10

\bibitem[{Habibi {et~al.}(2017)Habibi, Gillessen, Martins, Eisenhauer, Plewa,
  Pfuhl, George, Dexter, Waisberg, Ott, von Fellenberg, Baub{\"{o}}ck,
  Jimenez-Rosales, \& Genzel}]{Habibi2017}
Habibi, M., Gillessen, S., Martins, F., {et~al.} 2017, The Astrophysical
  Journal, 847, 120

\bibitem[{Hora {et~al.}(2014)Hora, Witzel, Ashby, Becklin, Carey, Fazio, Ghez,
  Ingalls, Meyer, Morris, Smith, \& Willner}]{Hora2014}
Hora, J.~L., Witzel, G., Ashby, M.~L., {et~al.} 2014, Astrophysical Journal,
  793, 120

\bibitem[{K{\"{u}}nsch(1989)}]{Kunsch1989}
K{\"{u}}nsch, H.~R. 1989, The Annals of Statistics, 17, 1217

\bibitem[{Li {et~al.}(2015)Li, Yuan, Yuan, Wang, Chen, Neilsen, Fang, Zhang, \&
  Dexter}]{Li2015}
Li, Y.~P., Yuan, F., Yuan, Q., {et~al.} 2015, The Astrophysical Journal, 810,
  19

\bibitem[{Liu {et~al.}(2016)Liu, Wright, Zhao, Brinkerink, {T. P. Ho}, Mills,
  Mart{\'{i}}n, Falcke, Matsushita, \& Mart{\'{i}}-Vidal}]{Liu2016}
Liu, H.~B., Wright, M. C.~H., Zhao, J.-H., {et~al.} 2016, Astronomy {\&}
  Astrophysics, 593, A107

\bibitem[{Mao {et~al.}(2016)Mao, Dexter, \& Quataert}]{Mao2016}
Mao, S.~A., Dexter, J., \& Quataert, E. 2016, Monthly Notices of the Royal
  Astronomical Society, 466, 4307

\bibitem[{Perrin \& Woillez(2019)}]{Perrin2019x}
Perrin, G. \& Woillez, J. 2019, Astronomy and Astrophysics, 625, 48

\bibitem[{Ponti {et~al.}(2017)Ponti, George, Scaringi, Zhang, Jin, Dexter,
  Terrier, Clavel, Degenaar, Eisenhauer, Genzel, Gillessen, Goldwurm, Habibi,
  Haggard, Hailey, Harrison, Merloni, Mori, Nandra, Ott, Pfuhl, Plewa, \&
  Waisberg}]{Ponti2017}
Ponti, G., George, E., Scaringi, S., {et~al.} 2017, Monthly Notices of the
  Royal Astronomical Society, 468, 2447

\bibitem[{Ressler {et~al.}(2017)Ressler, Tchekhovskoy, Quataert, \&
  Gammie}]{Ressler2017}
Ressler, S.~M., Tchekhovskoy, A., Quataert, E., \& Gammie, C.~F. 2017, Monthly
  Notices of the Royal Astronomical Society, 467, 3604

\bibitem[{Scargle(2020)}]{Scargle2020}
Scargle, J.~D. 2020, {Studies in Astronomical Time Series Analysis: VII. An
  Enquiry Concerning Non-Linearity, the RMS-Mean Flux Relation, and log-Normal
  Flux Distributions}, Tech. rep.

\bibitem[{Sch{\"{o}}del {et~al.}(2010)Sch{\"{o}}del, Najarro, Muzic, \&
  Eckart}]{Schodel2010}
Sch{\"{o}}del, R., Najarro, F., Muzic, K., \& Eckart, A. 2010, Astronomy and
  Astrophysics, 511, A18

\bibitem[{Scott(2015)}]{Scott2015}
Scott, D.~W. 2015, {Multivariate density estimation: Theory, practice, and
  visualization: Second edition}, 2nd edn. (John Wiley {\&} Sons, Inc.), 1--360

\bibitem[{Stone {et~al.}(2016)Stone, Marrone, Dowell, Schulz, Heinke, \&
  Yusef-Zadeh}]{Stone2016}
Stone, J.~M., Marrone, D.~P., Dowell, C.~D., {et~al.} 2016, The Astrophysical
  Journal, 825, 32

\bibitem[{Uttley {et~al.}(2005)Uttley, McHardy, \& Vaughan}]{Uttley2005}
Uttley, P., McHardy, I.~M., \& Vaughan, S. 2005, Monthly Notices of the Royal
  Astronomical Society, 359, 345

\bibitem[{Vaughan {et~al.}(2003)Vaughan, Edelson, Warwick, \&
  Uttley}]{Vaughan2003}
Vaughan, S., Edelson, R., Warwick, R.~S., \& Uttley, P. 2003, Monthly Notices
  of the Royal Astronomical Society, 345, 1271

\bibitem[{von Fellenberg {et~al.}(2018)von Fellenberg, Gillessen,
  Graci{\'{a}}-Carpio, Fritz, Dexter, Baub{\"{o}}ck, Ponti, Gao, Habibi, Plewa,
  Pfuhl, Jimenez-Rosales, Waisberg, Widmann, Ott, Eisenhauer, \&
  Genzel}]{VonFellenberg2018}
von Fellenberg, S.~D., Gillessen, S., Graci{\'{a}}-Carpio, J., {et~al.} 2018,
  The Astrophysical Journal, 862, 129

\bibitem[{Waisberg(2019)}]{ReisWaisberg2019}
Waisberg, I. 2019, {Optical Interferometry of Compact Objects: Testing General
  Relativity and the Extremes of Accretion}, Tech. rep.

\bibitem[{Witzel {et~al.}(2012)Witzel, Eckart, Bremer, Zamaninasab,
  Shahzamanian, Valencia-S., Sch{\"{o}}del, Karas, Lenzen, Marchili, Sabha,
  Garcia-Marin, Buchholz, Kunneriath, \& Straubmeier}]{Witzel2012}
Witzel, G., Eckart, A., Bremer, M., {et~al.} 2012, Astrophysical Journal,
  Supplement Series, 203, 18

\bibitem[{Witzel {et~al.}(2018)Witzel, Martinez, Hora, Willner, Morris, Gammie,
  Becklin, Ashby, Baganoff, Carey, Do, Fazio, Ghez, Glaccum, Haggard,
  Herrero-Illana, Ingalls, Narayan, \& Smith}]{Witzel2018}
Witzel, G., Martinez, G., Hora, J., {et~al.} 2018, The Astrophysical Journal,
  863, 15

\bibitem[{Yuan {et~al.}(2003)Yuan, Quataert, \& Narayan}]{Yuan2003}
Yuan, F., Quataert, E., \& Narayan, R. 2003, The Astrophysical Journal, 598,
  301

\end{thebibliography}
\bibliographystyle{aa}

\end{document}